\documentclass{article}

\usepackage{microtype}
\usepackage{graphicx}
\usepackage{subcaption}
\usepackage{booktabs} %
\usepackage{multirow}

\usepackage{hyperref}

\usepackage[preprint]{icml2026}

\usepackage{amsmath}
\usepackage{amssymb}
\usepackage{mathtools}
\usepackage{amsthm}
\usepackage{pifont}
\usepackage{makecell}

\usepackage[capitalize,noabbrev]{cleveref}

\theoremstyle{plain}

\theoremstyle{definition}

\theoremstyle{remark}

\usepackage[textsize=tiny]{todonotes}

\icmltitlerunning{Reasoning-Augmented Representations for Multimodal Retrieval}

\begin{document}

\twocolumn[
  \icmltitle{Reasoning-Augmented Representations for Multimodal Retrieval}

  \icmlsetsymbol{equal}{*}

  \begin{icmlauthorlist}
    \icmlauthor{Jianrui Zhang}{equal,yyy}
    \icmlauthor{Anirudh Sundara Rajan}{equal,yyy}
    \icmlauthor{Brandon Han}{yyy}\\
    \icmlauthor{Soochahn Lee}{zzz}
    \icmlauthor{Sukanta Ganguly}{xxx}
    \icmlauthor{Yong Jae Lee}{yyy}
  \end{icmlauthorlist}

  \icmlaffiliation{yyy}{University of Wisconsin-Madison}
  \icmlaffiliation{zzz}{Kookmin University}
  \icmlaffiliation{xxx}{NetApp, Inc}

  \icmlcorrespondingauthor{Jianrui Zhang}{harrisz@cs.wisc.edu}
  \icmlcorrespondingauthor{Anirudh Sundara Rajan}{asundararaj2@wisc.edu}

  \icmlkeywords{Machine Learning, ICML}

  \vskip 0.3in
]

\printAffiliationsAndNotice{\icmlEqualContribution}

\begin{abstract}
Universal Multimodal Retrieval (UMR) seeks any-to-any search across text and vision, yet modern embedding models remain brittle when queries require latent reasoning (e.g., resolving underspecified references or matching compositional constraints).
We argue this brittleness is often \emph{data-induced}: when images carry ``silent'' evidence and queries leave key semantics implicit, a single embedding pass must both reason and compress, encouraging spurious feature matching.
We propose a data-centric framework that decouples these roles by \emph{externalizing} reasoning before retrieval.
Using a strong Vision--Language Model, we make implicit semantics explicit by densely captioning visual evidence in corpus entries, resolving ambiguous multimodal references in queries, and rewriting verbose instructions into concise retrieval constraints.
Inference-time enhancement alone is insufficient; the retriever must be trained on these semantically dense representations to avoid distribution shift and fully exploit the added signal.
Across M-BEIR, our reasoning-augmented training method yields consistent gains over strong baselines, with ablations showing that corpus enhancement chiefly benefits knowledge-intensive queries while query enhancement is critical for compositional modification requests. We publicly release our code at~\url{https://github.com/AugmentedRetrieval/ReasoningAugmentedRetrieval}.
\end{abstract}

\begin{figure}
\centering
\includegraphics[width=.92\linewidth]{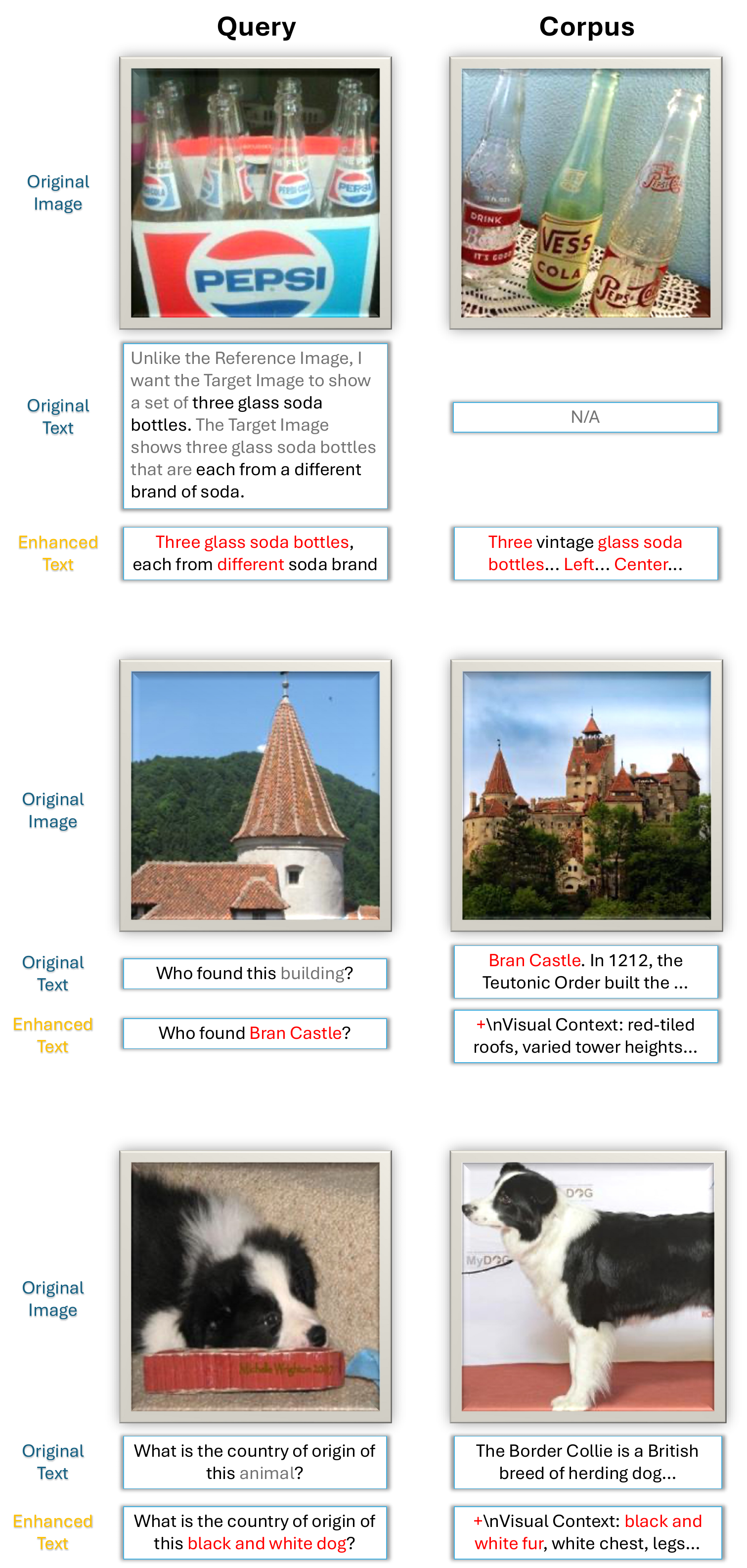}
\caption{
Overview of our reasoning-augmented multimodal representation. We use a strong VLM to (i) inject descriptive text into image-only inputs and (ii) refine captions for image--text pairs, making key visual semantics explicit. This externalizes implicit reasoning, allowing the retriever to focus on compression into robust embeddings rather than on-the-fly visual inference. Red highlights denote the matching components on both the query and corpus sides with the help of VLM-generated enhancements.}
\label{fig: teaser}
\end{figure}

\section{Introduction}

Universal Multimodal Retrieval (UMR) aims to support information seeking across arbitrary combinations of modalities---text-to-image, image-to-text, and image-to-image---within a single retrieval system. A dominant paradigm is to learn a shared embedding space in which semantically corresponding inputs are nearby. Large-scale contrastive pretraining has yielded strong multimodal representations: CLIP \citep{radford2021learningtransferablevisualmodels}, SigLIP \citep{zhai2023sigmoidlosslanguageimage}, and ImageBind \citep{girdhar2023imagebind} use modality-specific encoders and achieve impressive cross-modal retrieval performance. However, their designs are fundamentally strained by mixed-modality inputs (e.g., an image paired with a textual qualifier), typically resorting to ad hoc fusion strategies \citep{wei2023uniirtrainingbenchmarkinguniversal} that can be brittle. An alternative is to leverage multimodal large language models (MLLMs) \citep{liu2023visual} as embedding models, which can natively ingest and jointly interpret image--text inputs; recent work has demonstrated the promise of this direction \citep{jiang2024e5vuniversalembeddingsmultimodal,jiang2025vlm2vectrainingvisionlanguagemodels,liu2024lamralargemultimodalmodel}.

Despite this progress, recent benchmarks such as M-BEIR \citep{wei2023uniirtrainingbenchmarkinguniversal} highlight a persistent gap between retrieval and \emph{information seeking}. Many queries require nontrivial interpretation before they can be matched to a corpus entry. For example, given an image of an unfamiliar building and the query ``Who designed this?'', a retriever must implicitly (i) recognize or disambiguate the depicted structure, (ii) infer that the desired attribute is the \emph{architect}, and (iii) connect that attribute to the appropriate passage or entity in the corpus. When these intermediate steps remain implicit, existing models frequently fall back on superficial cues---matching sky color, viewpoint, or low-level texture---rather than the underlying intent, leading to failures driven by \textbf{spurious correlations}. These failure modes are not unique to any one architecture; they arise whenever the representation must simultaneously infer \emph{what to match} and compress the input into a single vector.

This observation motivates a simple but consequential hypothesis: expecting a single embedding forward pass to perform both \emph{reasoning} (deciding which latent aspects of the input are relevant) and \emph{compression} (mapping the result into a compact vector) is often too demanding, especially for underspecified queries. Queries in UMR commonly contain indirect references (``this'', ``the one on the left''), missing attributes (``the designer'' vs.\ ``the architect''), or verbose instructions whose core constraints are buried in irrelevant text. When the semantics are not made explicit, the embedding model is forced to approximate the missing reasoning with heuristics, which amplifies spurious matching. We therefore propose to \emph{decouple reasoning from compression} by explicitly materializing the latent reasoning steps as text annotations and then training the retriever to operate on this \emph{dense semantic} form.

Concretely, we introduce a \emph{reasoning-augmented enhancement} pipeline that shifts interpretive burden away from the embedding model and into a separate, high-capacity Vision--Language Model (VLM). The VLM is used to (i) caption salient visual details that are otherwise implicit, (ii) resolve ambiguous references in multimodal queries, and (iii) rewrite verbose instructions into concise, retrieval-oriented constraints. In effect, the pipeline \emph{externalizes} the hidden chain of reasoning that a retriever would otherwise need to perform internally. This turns difficult ``reasoning-then-retrieve'' problems into \emph{explicit semantic matching}: once the relevant attributes and referents are stated plainly, a standard embedding model can focus on faithful compression and similarity search rather than guessing what the query meant.

A key advantage of this approach is practicality. Corpus-side enhancement is a one-time offline cost, and query-side enhancement can be performed efficiently at test time with modest overhead. More importantly, we find that the benefit is not merely an inference-time artifact of higher-quality text. Training retrieval backbones on the enhanced distribution yields consistent and substantial improvements over strong baselines across M-BEIR \citep{wei2023uniirtrainingbenchmarkinguniversal}. Crucially, our results indicate that models must be \emph{re-aligned} to operate on these explicit, semantically dense inputs; simply enhancing queries at inference time is insufficient to eliminate spurious feature matching. Together, these findings suggest that robust UMR is less about ever more expressive embedding architectures and more about \emph{making the semantics matchable}: by converting implicit reasoning into explicit, verifiable descriptions, we enable retrieval models to generalize beyond shallow correlations and better satisfy real information-seeking intent.

\section{Related Work}

\paragraph{Foundational multimodal retrieval.}
Early multimodal retrieval systems largely relied on dual-encoder architectures trained with contrastive objectives to align image and text representations in a shared embedding space.
CLIP \citep{radford2021learningtransferablevisualmodels} and SigLIP \citep{zhai2023sigmoidlosslanguageimage} demonstrated that large-scale image--text contrastive pretraining enables strong zero-shot transfer and cross-modal matching.
Building on this paradigm, UniIR \citep{wei2023uniirtrainingbenchmarkinguniversal} introduced an instruction-guided retriever for \emph{universal} multimodal retrieval and established the M-BEIR benchmark, which emphasizes realistic information-seeking queries across heterogeneous tasks.
However, dual encoders typically process each modality independently and compress inputs into a single vector, which can be brittle when queries require \emph{joint interpretation} of mixed-modal inputs or latent reasoning (e.g., resolving references such as ``this'' or inferring the intended attribute in ``Who designed this?'').
This brittleness is increasingly exposed by M-BEIR-style settings, where shallow correlations can dominate when the underlying intent is underspecified.

\vspace{-5pt}
\paragraph{VLM-based universal embeddings.}
The rise of large vision--language models (VLMs) has shifted retrieval toward embedding models that can natively consume multimodal inputs and perform richer cross-modal interactions before compression.
E5-V \citep{jiang2024e5vuniversalembeddingsmultimodal} adapts VLM backbones for universal embeddings and argues that strong transfer can be obtained even with predominantly text-pair supervision.
In contrast, VLM2Vec \citep{jiang2025vlm2vectrainingvisionlanguagemodels} demonstrates benefits from explicitly multimodal supervision. Subsequent work has focused on training recipes and architectural adaptations that improve universal embedding quality, including specialized pooling or embedding tokens \citep{liu2024lamralargemultimodalmodel}, hard-negative strategies and reranker distillation \citep{li2025umarvelunveilingkeyfactors}, and multi-stage pipelines that achieve strong performance on newer benchmarks \citep{li2026qwen3vlembeddingqwen3vlrerankerunifiedframework}.
These advances primarily improve \emph{how} embeddings are learned (architecture, losses, distillation, mining).
In contrast, our work targets \emph{what} is being embedded: we identify a data/representation bottleneck where queries and corpus entries lack sufficient semantic explicitness, forcing embedding models to simultaneously infer intent and compress, which encourages spurious matching.
Rather than modifying the retriever architecture, we use a VLM to externalize latent reasoning into explicit textual form, and show that training on this semantically denser distribution yields consistent gains.

\vspace{-5pt}
\paragraph{Data-centric and task-specialized retrieval.}
A complementary line of research improves retrieval by tailoring representations and interaction mechanisms to specific modalities or by altering the training data.
For document-centric visual retrieval, ColPali \citep{faysse2025colpaliefficientdocumentretrieval} adopts late interaction \citep{khattab2020colbert} to reduce reliance on brittle OCR pipelines and better capture token-level evidence.
Extensions such as VLM2Vec-v2 \citep{meng2025vlm2vecv2advancingmultimodalembedding} broaden universal embeddings to additional modalities (e.g., video) and domains (e.g., visual documents).
On the data side, GME \citep{zhang2025gmeimprovinguniversalmultimodal} synthesizes large-scale fused-modal training pairs to mitigate modality imbalance and enrich supervision.
While such approaches increase coverage by creating \emph{new} training data or designing modality-specific mechanisms, our approach is orthogonal and lightweight: we refine the \emph{semantic density} of the existing benchmark inputs by resolving ambiguities, explicitly captioning salient visual evidence, and rewriting verbose instructions into retrieval-oriented constraints.
This shifts implicit reasoning from the embedding model to an external VLM, effectively converting ``reason-then-retrieve'' into explicit semantic matching, and complements prior improvements in architecture and training strategy.

\section{Method}
\label{sec:method}

Our goal is to improve universal multimodal retrieval by \emph{decoupling reasoning from compression}. In standard embedding-based retrieval, a single model must (i) infer the intended semantics of an underspecified multimodal input (e.g., resolve textual references to objects in an image or identify described visual attributes) and (ii) compress the result into a compact vector for nearest-neighbor search. If these reasoning steps are implicitly learned, embeddings can overfit to shallow cues and spurious correlations.
We instead \emph{explicitly externalize} the intended semantics of inputs by using a strong VLM (in our case, Qwen3-VL-8B~\citep{bai2025qwen3vltechnicalreport}) to produce semantically dense, self-contained textual information that allows embedding models to focus on data compression.

We implement this idea through a \textbf{reasoning-augmented enhancement pipeline} with the following two components:
(1) \emph{corpus enhancement}, which ``unsilences'' visual evidence in database items by adding explicit textual descriptions of visual information; and
(2) \emph{query enhancement}, which resolves ambiguous references and rewrites verbose instructions into retrieval-oriented constraints.
The enhanced data is then used to train (or fine-tune) a standard retrieval backbone, aligning the embedding space to this semantically explicit distribution.

\subsection{Problem Setup and Notation}
\label{sec:setup}
Let a retrieval corpus be $\mathcal{D}=\{d_i\}_{i=1}^N$, where each item $d_i$ may contain text $t_i$, an image $v_i$, or both.
A query $q$ similarly may contain text $t_q$, an image $v_q$, or both.
An embedding retriever parameterized by $\theta$ maps an input to a vector representation:
\[
\mathbf{z}_q = f_\theta(q), \qquad \mathbf{z}_i = f_\theta(d_i),
\]
and retrieves by similarity search between $\mathbf{z}_q$ and $\mathbf{z}_i$.

Our method introduces an \emph{enhancer} $E$ (a VLM) that transforms inputs to have more explicit textual forms:
\[
\tilde{d}_i = E(d_i), \qquad \tilde{q} = E(q),
\]
where $\tilde{d}_i$ and $\tilde{q}$ are semantically denser, self-contained representations. The retriever is trained on these enhanced forms, i.e.,
\[
\mathbf{z}_q = f_\theta(\tilde{q}), \qquad \mathbf{z}_i = f_\theta(\tilde{d}_i).
\]
Intuitively, $E$ performs the ``interpretation'' step (e.g., captioning, reference resolution, constraint extraction), while $f_\theta$ focuses on stable compression for retrieval.

\subsection{Corpus Enhancement}
\label{sec:corpus_enhancement}
Real-world multimodal corpora typically contain three entry types:
(I) text-only, (II) image-only, and (III) image--text pairs. Our enhancement targets \emph{missing visual} semantics; thus Category~I is left unchanged.

\vspace{-5pt}
\paragraph{Category II: image-only entries (unsilencing visual evidence).}
For image-only items, the decisive evidence is often entirely visual, making them ``silent'' to a text-centric embedding model.
Therefore, to effectively extract the image semantics and enrich the dataset item, we generate a \emph{dense visual caption} that enumerates salient objects, attributes, relations, and distinctive cues.
We prompt the VLM to produce a \emph{keyword-rich} description (about 100 words) rather than fluent prose.
This design prioritizes \emph{lexical coverage} of discriminative attributes (e.g., ``arched bridge,'' ``gothic facade,'' ``two people holding surfboards''), which empirically is more helpful for embedding-based matching than stylistic verbosity.
Dense captions also reduce ambiguity by making implicit evidence explicit and directly matchable.

\vspace{-5pt}
\paragraph{Category III: image--text entries (preserving metadata while adding visual grounding).}
In image--text entries, the accompanying text often only provide contextual information like encyclopedic content. Textual descriptions of the visual data are either high-level at best. To enrich these entries, we \emph{append} an image caption to the original query text, i.e.,
\begin{center}
``\{original text\}\textbackslash nVisual Context: \{caption\}''.
\end{center}
Appending our generated caption to the existing data enjoys two benefits over either entirely replacing the original text or interleaving within it. Firstly, the original text may include critical metadata (e.g., names, dates, definitions, or narrative content) that distinguishes our corpus item. Altering or removing this data would have a significant negative impact, especially in cases where multiple corpus items have similar images. Secondly, it makes the augmentation robust to moderate caption noise, since the retriever can learn to rely on whichever span (i.e., original or caption) provides the most reliable evidence for the task.

\subsection{Query Enhancement}
\label{sec:query_enhancement}
Queries follow the same three types as corpus entries. We do not modify Category~I (text-only) queries.
For the remaining types, the objective is to convert underspecified multimodal queries into \emph{self-contained} text that exposes the intended semantics.

\vspace{-5pt}
\paragraph{Category II: image-only queries.}
For image-only queries, we generate a dense caption using the same schema as for corpus images, but with a tighter length budget (about 50 words).
We present captions that should capture the primary discriminative evidence while avoiding overly granular details that can introduce noise and dilute the query intent.

\vspace{-5pt}
\paragraph{Category III: image--text queries.}
Mixed-modal queries commonly fall into two formats, which require different forms of externalized reasoning:
(a) \emph{QA-style questions} that require reference resolution and intent completion, and
(b) \emph{modification requests} that require constraint extraction and target specification.

\subsubsection{QA-style queries}
QA-style queries often include ambiguous deictic references (e.g., ``When was \emph{this animal} discovered?'') whose referent is only identifiable from the image.
To retrieve the correct document, a system must first resolve the referent and make the question explicit.
We prompt the VLM to rewrite such queries into explicit text by:
(i) identifying the referent when it is recognizable (e.g., ``giant panda'', ``Eiffel Tower''), and
(ii) substituting vague phrases (``this animal'', ``this building'') with the resolved entity name.
When a canonical name is uncertain, the VLM emits a concise visual descriptor (e.g., ``a red noodle soup with shrimp''; ``a black-and-yellow wasp'').
The result is a self-contained query that can be matched via standard text-based semantics, reducing the need for the retriever to perform implicit visual reasoning.

\subsubsection{Modification requests}
Modification-style queries specify a desired transformation of a reference image (e.g., given an image of a dog in a field, ``change to cat''), where the retrieval target should satisfy the modified constraints (e.g., ``a cat running in a field'').
These requests are frequently verbose, include superfluous phrasing, and bury the actionable constraints.
We prompt the VLM to distill the request into a concise, keyword-based constraint set (e.g., ``Unlike the reference image, I want the target image to have shorter hair and more dogs'' $\rightarrow$ ``shorter hair; more dogs'').
This representation is well-suited to embedding retrieval: it foregrounds the discriminative attributes the target should satisfy and improves compositional matching.

Crucially, when rewriting modification requests, we exclude the reference image from the VLM input.
VLMs trained with dense captioning priors exhibit a strong tendency to describe the visible image, even when the instruction concerns a \emph{target} that differs from the reference.
Including the image can cause \emph{contextual contamination}, where the rewritten query is biased toward properties of the reference rather than the intended modification.
Conditioning only on the instruction text encourages faithful constraint extraction and yields more consistent retrieval-oriented rewrites.

\subsection{Training on Enhanced Representations}
\label{sec:training_on_enhanced}
The enhancement pipeline changes the \emph{semantic density} and surface statistics of both queries and corpus entries.
Accordingly, we train the retrieval backbone on enhanced pairs so that its embedding space is aligned with the explicit semantics produced by $E$.
This training step is conceptually simple---it uses standard retrieval objectives (e.g., contrastive learning) on $(\tilde{q}, \tilde{d}^+)$ pairs---but is essential for ensuring that the model leverages the enhanced signal rather than defaulting to shallow correlations learned from the original, underspecified distribution.

\begin{table*}[t]
\centering
\caption{Recall Scores Comparison (Baseline vs. Ours). Each task is named after the original benchmark name dash the task number. Tasks are grouped by retrieval category: General, Knowledge, and Composed. Each task type is labelled by query type ($i$ means image and $t$ means text) $\to$ corpus type. Improvements using Ours are bolded. We gray out the results for FashionIQ and Fashion200k due to unreliable ground-truth.}
\label{tab:recall_scores}
\begin{tabular}{l|l|cccc|cccc}
\toprule
\multirow{2}{*}{\textbf{Task}} & \multirow{2}{*}{\textbf{Type}} & \multicolumn{4}{c|}{\textbf{ Baseline \tiny{(LamRA-Ret~\citep{liu2024lamralargemultimodalmodel})}}} & \multicolumn{4}{c}{\textbf{Ours}} \\
 & & \textbf{R@1} & \textbf{R@5} & \textbf{R@10} & \textbf{R@50} & \textbf{R@1} & \textbf{R@5} & \textbf{R@10} & \textbf{R@50} \\
 \midrule
MSCOCO-0 & $q^t\to c^i$ & 54.19 & 79.36 & 87.18 & 97.96 & \textbf{55.57} & \textbf{80.20} & \textbf{87.62} & \textbf{98.12} \\
MSCOCO-3 & $q^i\to c^t$ & 69.90 & 89.66 & 94.42 & 99.36 & \textbf{70.80} & \textbf{90.02} & 94.34 & \textbf{99.52} \\
VisNews-0 & $q^t\to c^i$ & 13.95 & 28.47 & 35.70 & 53.89 & \textbf{14.98} & \textbf{29.70} & \textbf{37.14} & \textbf{55.42} \\
VisNews-3 & $q^i\to c^t$ & 12.08 & 26.16 & 33.62 & 52.00 & \textbf{12.88} & \textbf{26.82} & \textbf{34.17} & \textbf{53.02} \\
WebQA-1 & $q^t\to c^t$ & 57.31 & 83.58 & 90.14 & 97.07 & \textbf{58.90} & \textbf{84.89} & \textbf{91.00} & \textbf{97.39} \\
WebQA-2 & $q^t\to c^i,c^t$ & 49.54 & 77.78 & 86.62 & 96.26 & \textbf{51.10} & \textbf{79.61} & \textbf{86.74} & \textbf{96.81} \\
InfoSeek-6 & $q^i,q^t\to c^t$ & 21.64 & 41.43 & 51.51 & 72.84 & \textbf{22.10} & \textbf{42.08} & 51.13 & 71.37 \\
InfoSeek-8 & $q^i,q^t\to c^i,c^t$ & 25.50 & 49.72 & 60.81 & 80.34 & \textbf{28.34} & \textbf{52.38} & \textbf{62.21} & \textbf{80.93} \\
OVEN-6 & $q^i,q^t\to c^t$ & 24.82 & 47.96 & 57.79 & 76.22 & \textbf{25.47} & \textbf{48.32} & \textbf{57.89} & 75.94 \\
OVEN-8 & $q^i,q^t\to c^i,c^t$ & 47.65 & 68.11 & 75.24 & 87.37 & 47.46 & \textbf{68.56} & 74.80 & 86.69 \\

EDIS-2 & $q^t\to c^i,c^t$ & 26.97 & 53.38 & 64.33 & 82.81 & \textbf{27.12} & 53.29 & 63.87 & \textbf{83.46} \\
Nights-4 & $q^i\to c^i$ & 8.40 & 30.85 & 49.01 & 92.88 & 8.02 & \textbf{31.56} & \textbf{50.57} & 92.74 \\
CIRR-7 & $q^i\to c^i$ & 21.41 & 49.35 & 60.77 & 82.81 & \textbf{22.71} & \textbf{49.40} & 59.93 & 82.73 \\
\color{gray} FashIQ-7 & \color{gray} $q^i,q^t\to c^i$ & \color{gray} 10.48 & \color{gray} 23.55 & \color{gray} 30.67 & \color{gray} 50.07 & \color{gray} 10.48 & \color{gray} 22.92 & \color{gray} 30.35 & \color{gray} 49.01 \\
\color{gray} Fash200k-3 & \color{gray} $q^i\to c^t$ & \color{gray} 4.91 & \color{gray} 13.11 & \color{gray} 18.82 & \color{gray} 37.98 & \color{gray} 4.89 & \color{gray} 13.07 & \color{gray} \textbf{18.88} & \color{gray} 37.90 \\
\bottomrule
\end{tabular}%
\end{table*}

\section{Experiments}\label{sec:experiments}

In this section, we conduct exhaustive experiments by training retrieval models on both the enhanced and unenhanced M-BEIR datasets~\citep{wei2023uniirtrainingbenchmarkinguniversal}. We use M-BEIR due to its vast coverage of multiple task types of multimodal retrieval. We first discuss our training recipe in \S~\ref{sec: train recipe}, then discuss our main results in \S~\ref{sec: main results}. We conduct ablation studies dissecting performance improvements in \S~\ref{sec: ablation 1} and justifying the necessity of training on our enhanced data in \S~\ref{sec: ablation 2}. Finally, we provide qualitative results to more directly demonstrate the superiority of our method in \S~\ref{sec: qual}. We also report results on composed image retrieval tasks from MVRB \cite{liu2025any} in Appendix \ref{app:mvrb} and report results of variants trained with hard negatives in Appendix~\ref{app: hard negs}.

\subsection{Training Recipe}
\label{sec: train recipe}
We adopt the training recipe from LamRA-Ret~\citep{liu2024lamralargemultimodalmodel}, which adapts a VLM for retrieval tasks by introducing a special embedding token, \texttt{<emb>}. Specifically, we prepend task-specific instructions and append the prompt ``Summarize the above into one word: \texttt{<emb>}'' to the input sequence and utilize the second-last hidden state corresponding to this token to compute contrastive losses between the query and the positive candidates from the corpus. We skip stage 1 text-only pretraining as we only focus on multimodal retrieval. We use Qwen3-VL-2B~\citep{bai2025qwen3vltechnicalreport}, an open-source state-of-the-art VLM chosen for its balance of performance and deployment efficiency, as our base model to finetune from for retrieval. Each model takes only 7 hours to train, and takes around 2-3 hours to evaluate on all reported benchmarks. 

For all query and corpus enhancements, we utilize Qwen3-VL-8B~\citep{bai2025qwen3vltechnicalreport}. Our methodology requires that both the corpus and query sets undergo enhancement prior to training. To facilitate this, we provide scripts and prompts enabling Qwen to generate enhanced queries and captions for the target database. Utilizing 8$\times$A100-80GB GPUs, we can complete the enhancement of the entire M-BEIR corpus and query set—a dataset exceeding 7 million entries—in approximately 8 hours. During inference, our code can easily be incorporated into the system as an online enhancement tool to refine user queries on the go with negligible overhead. All prompts utilized in this pipeline are provided in Appendix~\ref{app: prompts}.

\subsection{Main Results}
\label{sec: main results}

The comprehensive results of our enhanced retrieval pipeline are presented in Table~\ref{tab:recall_scores}. By systematically addressing the semantic gaps in both queries and corpus entries, our method achieves consistent improvements across the majority of M-BEIR benchmarks compared to the baseline LamRA implementation. 

A critical observation in our results is the substantial improvement in the WebQA and InfoSeek benchmarks (e.g., Infoseek-8 R@1 +2.84\%, WebQA-1 R@1 +1.59\%). Our query enhancement—resolving ambiguous terms like ``this animal'' into named entities (e.g., ``Great Panda'')—directly benefits InfoSeek and OVEN-6, where precise entity matching is paramount. We even observe a significant $\sim$1.6\% improvement on WebQA, a text-to-text retrieval task. It is important to highlight that our method \textit{does not} alter text-to-text retrieval pathways; we strictly preserve original text-only queries and corpus entries. Despite this, we observe outsized gains in these tasks. This confirms that the primary bottleneck in previous baselines was not the textual modality itself, but the disconnect between textual queries and visual evidence. By augmenting the image-text corpus with our generated ``Visual Context", the modality gap between image and text is further reduced during training. We hypothesize that this has indirectly caused the improvement in the quality of text-only embeddings. 

For standard cross-modal tasks such as MSCOCO and VisualNews, our corpus enhancement strategy yields notable gains (e.g., VisNews-0 R@1 +1.03\%). In these datasets, the raw images often contain visual details that are not captured in the original, often noisy, alt-text or captions. Our dense captioning pipeline extracts these latent features into keyword-rich text, providing the retrieval model with a higher-resolution semantic signal during the contrastive learning phase.

In the CIRR benchmark, which tests the model's ability to modify a visual query based on text instructions (e.g., ``change dog to cat''), we achieve a clear improvement over the baseline (R@1 +1.3\%). This validates our ``Change Request" query enhancement strategy. By distilling verbose natural language commands (e.g., ``I would like the new image to show...'') into concise, keyword-driven instructions, we reduce the cognitive load on the retrieval model, preventing it from being distracted by irrelevant linguistic noise.

\begin{figure}[t]
\centering
\includegraphics[width=\linewidth]{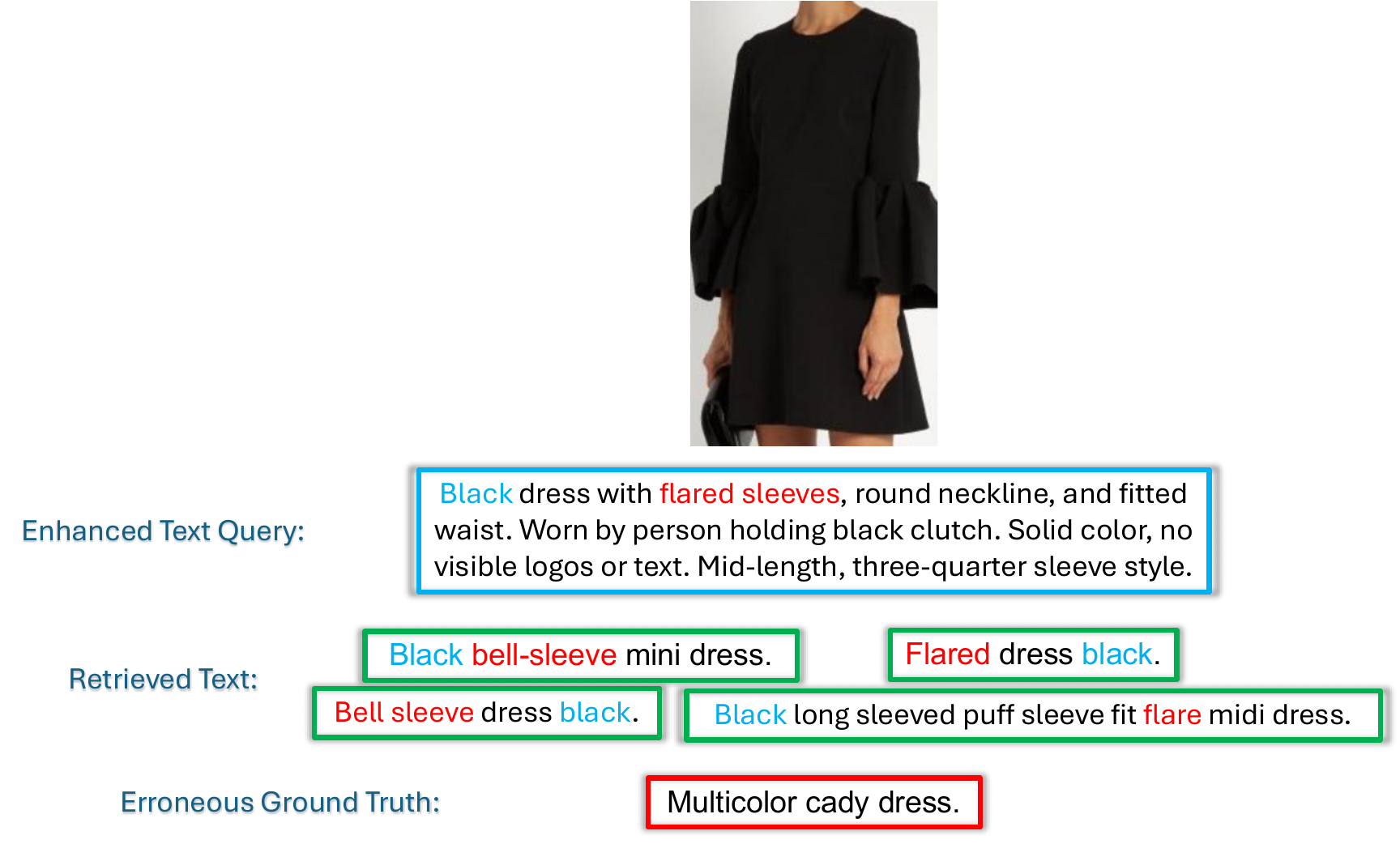}
\caption{In this entry of Fashion200K, the query asks for a description matching the image, which Qwen accurately captioned with keywords such as ``black dress" and ``flared sleeves". This allowed the retrieval of multiple correct results, indicating how the one-ground-truth design is flawed. Furthermore, the ground truth's ``multicolored" description is simply untrue, reinforcing our observation over these benchmarks' low quality nature.}
\label{fig: fashion issue}
\end{figure}

We explicitly gray out the results for \textit{FashionIQ} and \textit{Fashion200k} in Table~\ref{tab:recall_scores}. These benchmarks present a unique evaluation challenge: while the corpus contains tens of thousands of items, the ground truth for a given query is often restricted to a single specific image ID. In the fashion domain, however, hundreds of items may share identical visual attributes (e.g., ``black leather jacket''). Our enhanced model frequently retrieves these visually correct candidates, but because they do not match the specific ground truth ID, they are penalized as false negatives. Thus, the flat or slightly regressed scores in these rows reflect a limitation of the evaluation metric rather than a degradation in model capability. An exceptional error case from \textit{Fashion200k-3} can be seen in Figure~\ref{fig: fashion issue}, where the enhanced query accurately describes the clothing in the image, allowing the retrieval of multiple results of similar meanings, while the ground truth is erroneous.

\begin{table}[t]
\centering
\caption{Ablation of component contributions on Recall@1. \textbf{Q-Only}: Training with enhanced queries but original corpus. \textbf{C-Only}: Training with enhanced corpus but original queries. We observe that corpus enhancement drives knowledge tasks, while query enhancement is critical for modification tasks. Best performances per row are bolded.}
\label{tab:ablation_per_task}
\resizebox{\linewidth}{!}{%
\begin{tabular}{l|c|cc|c}
\toprule
\textbf{Task} & \textbf{Baseline} & \textbf{Q-Only} & \textbf{C-Only} & \textbf{Full} \\
\midrule
MSCOCO-0 & 54.19 & 54.77 & 54.89 & \textbf{55.57} \\
MSCOCO-3 & 69.90 & \textbf{71.16} & 70.20 & 70.80 \\
VisNews-0 & 13.95 & 13.80 & 14.69 & \textbf{14.98} \\
VisNews-3 & 12.08 & 12.78 & 12.47 & \textbf{12.88} \\
WebQA-1 & 57.31 & 56.17 & 56.17 & \textbf{58.90} \\
WebQA-2 & 49.54 & 50.85 & 49.14 & \textbf{51.10} \\
InfoSeek-6 & 21.64 & 21.61 & 19.89 & \textbf{22.10} \\
InfoSeek-8 & 25.50 & \textbf{30.37} & 25.98 & 28.34 \\
OVEN-6 & 24.82 & 25.42 & 24.62 & \textbf{25.47} \\
OVEN-8 & 47.65 & 47.43 & 47.55 & 47.46 \\
EDIS-2 & 26.97 & 26.81 & \textbf{27.89} & 27.12 \\
Nights-4 & 8.40 & 8.49 & 7.36 & 8.02 \\
CIRR-7 & 21.41 & \textbf{22.76} & 22.73 & 22.71 \\
\color{gray} FashIQ-7 & \color{gray} 10.48 & \color{gray} 10.21 & \color{gray} 10.86 & \color{gray} 10.48 \\
\color{gray} Fash200k-3 & \color{gray} 4.91 & \color{gray} 5.03 & \color{gray} 4.91 & \color{gray} 4.89 \\
\midrule
\textbf{Average} & 29.92 & 30.51 & 29.96 & \textbf{30.72} \\
\bottomrule
\end{tabular}%
}
\end{table}

\subsection{Ablation Study 1: Disentangling Enhancements}
\label{sec: ablation 1}

To disentangle the contributions of our pipeline, we evaluate the retrieval performance (R@1) across four settings: the unenhanced baseline, query enhancement only (\textbf{Q-Only}), corpus enhancement only (\textbf{C-Only}), and the full pipeline (\textbf{Full}). The results are summarized in Table~\ref{tab:ablation_per_task}.

\vspace{-5pt}
\paragraph{Synergy of Components}
The results demonstrate that our full pipeline achieves the highest average recall (30.72), confirming that query and corpus enhancements are complementary. While individual modules provide gains, their combination is essential for \textit{MSCOCO-0}, \textit{VisNews-0/3}, \textit{WebQA-2}, and \textit{InfoSeek-6}. This suggests that complex QA tasks require both a clearer question (from Q-Enhancement) and a semantically richer target (from C-Enhancement) to bridge the modality gap.

\vspace{-5pt}
\paragraph{Impact of Query Enhancement}
Surprisingly, Query Enhancement alone (\textbf{Q-Only}) drives the majority of the performance gains (Avg: 30.51), significantly outperforming Corpus Enhancement alone (Avg: 29.96). This gap is mostly contributed by \textit{InfoSeek-8}, where Q-Only yields a remarkable improvement (+4.87) over the baseline, even surpassing the Full model by roughly 2 points (30.37 vs. 28.34). We hypothesize that for entity-centric questions, simply resolving ambiguous terms (e.g., replacing ``this building'' with ``The Empire State Building'') is more effective than enriching the corpus, whose much longer text entry length (composed of both the original Wikipedia entry and the dense caption) makes exact entity matching more difficult.

\vspace{-5pt}
\paragraph{Task-Specific Analysis}
In text-to-text retrieval task \textit{WebQA-1}, we realize that both Q-only and C-only reduces performance, meaning that without enhancing both sides of the retrieval, we do not achieve the indirect positive effect on text-only retrieval.
In standard cross-modal tasks like \textit{MSCOCO-3} (Image$\to$Text), Q-Only achieves the highest individual score (71.16), suggesting that dense captioning of the visual query is highly effective for matching against standard text captions.
For \textit{CIRR}, both modules contribute equally to the improvement (both recalls being $\sim$22.7), indicating that either clarifying the modification instruction or enriching the target image with text description is sufficient to solve Image-Text$\to$Image retrieval tasks.
An exception is that to the seemingly fact that Q-Only is generally superior, \textit{EDIS-2} favors C-Only (27.89), likely because event-centric retrieval benefits more from detailed contextual descriptions in the corpus than from query reformulation.

\subsection{Ablation Study 2: The Necessity of Training}
\label{sec: ablation 2}

A natural question arises: do the performance gains stem solely from the information-rich content of our enhanced queries and corpus, or is the training process necessary to utilize this information? To investigate this, we conducted an ``Inference-Only'' experiment where we utilized the baseline model (trained on the original M-BEIR) but performed retrieval using our enhanced queries and corpus.

\begin{table}[ht]
\centering
\caption{Analysis of Training Necessity using R@1. \textbf{Inference-Only} denotes using the unenhanced baseline model to retrieve using enhanced data. The performance drop indicates a significant distribution shift, confirming that the model must be explicitly trained to align with the dense semantic signals in our enhanced data.}
\label{tab:inference_only}
\resizebox{0.8\linewidth}{!}{%
\begin{tabular}{l|c|c|c}
\toprule
\textbf{Task} & \textbf{Baseline} & \textbf{Inf-Only} & \textbf{Full} \\
\midrule
MSCOCO-0 & 54.19 & 45.37 & \textbf{55.57} \\
MSCOCO-3 & 69.90 & 70.06 & \textbf{70.80} \\
VisNews-0 & 13.95 & 9.91 & \textbf{14.98} \\
VisNews-3 & 12.08 & 13.68 & \textbf{12.88} \\
WebQA-1 & 57.31 & 57.31 & \textbf{58.90} \\
WebQA-2 & 49.54 & 44.84 & \textbf{51.10} \\
InfoSeek-6 & 21.64 & 20.37 & \textbf{22.10} \\
InfoSeek-8 & 25.50 & 22.77 & \textbf{28.34} \\
OVEN-6 & 24.82 & 22.30 & \textbf{25.47} \\
OVEN-8 & 47.65 & 45.87 & 47.46 \\
EDIS-2 & 26.97 & 22.53 & \textbf{27.12} \\
Nights-4 & 8.40 & 5.99 & 8.02 \\
CIRR-7 & 21.41 & 17.41 & \textbf{22.71} \\
\color{gray} FashIQ-7 & \color{gray} 10.48 & \color{gray} 9.21 & \color{gray} 10.48 \\
\color{gray} Fash200k-3 & \color{gray} 4.91 & \color{gray} 4.87 & \color{gray} 4.89 \\
\midrule
\textbf{Average} & 29.92 & 27.50 & \textbf{30.72} \\
\bottomrule
\end{tabular}%
}
\end{table}

As shown in Table~\ref{tab:inference_only}, directly applying enhanced data to the baseline model results in performance degradation even when compared to the unenhanced baseline. We attribute this to a severe \textbf{distribution shift}. The baseline model, trained on sparse and often ambiguous text, fails to effectively process the dense, keyword-rich ``Visual Context" provided by our method, likely treating the additional tokens as noise rather than signal. This finding underscores that high-quality data alone is insufficient; the retrieval model must be explicitly aligned via training to effectively leverage the augmented semantic information.

\subsection{Comparison with Other Baselines}\label{baseline-comp}

\begin{table}[ht]
\centering
\caption{R@5 Scores Comparison (Incorporating Stronger Baselines). We include SigLIP, DSE~\citep{ma2024unifyingmultimodalretrievaldocument}, E5-V, and GME-2B. "Ours" demonstrates competitive performance, particularly against SigLIP on Knowledge tasks and DSE on General/Composed tasks. Best performance per row is \textbf{bolded} and second-best performance is \underline{underlined}.}
\label{tab:recall_scores_expanded}
\resizebox{\linewidth}{!}{%
\begin{tabular}{l|ccccc|c}
\toprule
\textbf{Task} & \textbf{CLIP} & \textbf{SigLIP} & \textbf{DSE} & \textbf{E5-V} & \textbf{GME-2B} & \textbf{Ours} \\
\midrule
MSCOCO-0 & 61.10 & \underline{75.70} & 74.62 & 52.38 & 71.82 & \textbf{80.20} \\
MSCOCO-3 & 79.00 & \underline{88.20} & 82.06 & 86.40 & 84.12 & \textbf{90.02} \\
VisNews-0 & \textbf{43.30} & 30.10 & 14.12 & 29.46 & \underline{38.85} & 29.70 \\
VisNews-3 & \textbf{41.30} & 30.80 & 8.74 & 29.54 & \underline{38.32} & 26.82 \\
Nights-4 & 26.10 & 28.90 & 27.36 & 27.92 & \underline{29.86} & \textbf{31.56} \\
WebQA-1 & 36.20 & 39.80 & 83.95 & \underline{89.94} & \textbf{95.19} & 84.89 \\
WebQA-2 & 45.10 & 43.50 & 66.99 & 49.62 & \textbf{83.15} & \underline{79.61} \\
InfoSeek-6 & 20.50 & 25.10 & 3.06 & 12.69 & \underline{39.06} & \textbf{42.08} \\
InfoSeek-8 & 26.40 & 27.40 & 5.96 & 39.69 & \underline{44.21} & \textbf{52.38} \\
OVEN-6 & 24.20 & 29.70 & 0.38 & 14.40 & \textbf{58.17} & \underline{48.32} \\
OVEN-8 & 38.80 & 41.70 & 0.39 & 54.46 & \textbf{75.98} & \underline{68.56} \\
EDIS-2 & 43.30 & 27.00 & 41.26 & 49.62 & \textbf{70.32} & \underline{53.29} \\
CIRR-7 & 13.20 & 22.70 & 36.52 & 13.19 & \underline{46.83} & \textbf{49.40} \\
\bottomrule
\end{tabular}%
}
\end{table}

In Table~\ref{tab:recall_scores_expanded}, we benchmark our approach against a diverse set of strong baselines, including dual-encoders (CLIP \citep{radford2021learningtransferablevisualmodels}, SigLIP \citep{zhai2023sigmoidlosslanguageimage}), dense retrievers (DSE \citep{ma2024unifying}), and recent VLM-based embedding models (E5-V \citep{jiang2024e5vuniversalembeddingsmultimodal}, GME-2B \citep{zhang2025gmeimprovinguniversalmultimodal}). 

Our data-centric pipeline achieves \textbf{state-of-the-art performance (R@5) on 6 out of the 13 reported tasks} and secures the second-best position on 4 others. This empirical evidence suggests that the "ceiling" of current retrieval models is often limited not by model capacity, but by the semantic sparsity of the training data.

It is critical to contextualize the performance of GME-2B against our method. GME-2B benefits from a massive training corpus of approximately 8 million image-text pairs, including a 1M subset that remains closed-source. In contrast, our model is trained strictly on the standard MBEIR dataset ($\sim$1M pairs). Despite this $8\times$ data disparity, our method outperforms GME-2B on  benchmarks such as \textit{MSCOCO}, \textit{InfoSeek}, and \textit{CIRR}. This demonstrates that our performance gains stem from the efficacy of our method rather than data scaling, offering a significantly more data-efficient solution.

The advantages of our reasoning-augmented data are most pronounced in knowledge-intensive tasks. On \textit{InfoSeek-8}, which requires complex entity linking and visual reasoning, our method achieves an R@5 of \textbf{52.38}, outperforming the strongest competitor (GME-2B) by over 8 points and nearly doubling the performance of SigLIP (27.40). This validates our hypothesis that resolving ambiguous coreferences (e.g., ``this lake'') into explicit entities during training enables standard models to perform reasoning tasks that typically stump visual-only encoders.

On the \textit{CIRR} modification task, we surpass both DSE (specialized for dense text) and GME-2B, achieving an R@5 of \textbf{49.40}. This indicates that our query enhancement strategy—distilling verbose instructions into keywords—is more effective for modification retrieval than simply scaling model size. Furthermore, on standard cross-modal tasks like \textit{MSCOCO}, we maintain a clear lead (90.02 on \textit{MSCOCO-3}), proving that our enhancements improve complex reasoning capabilities without sacrificing general zero-shot retrieval performance.

\begin{figure*}[t]
    \centering
    \includegraphics[width=.95\linewidth]{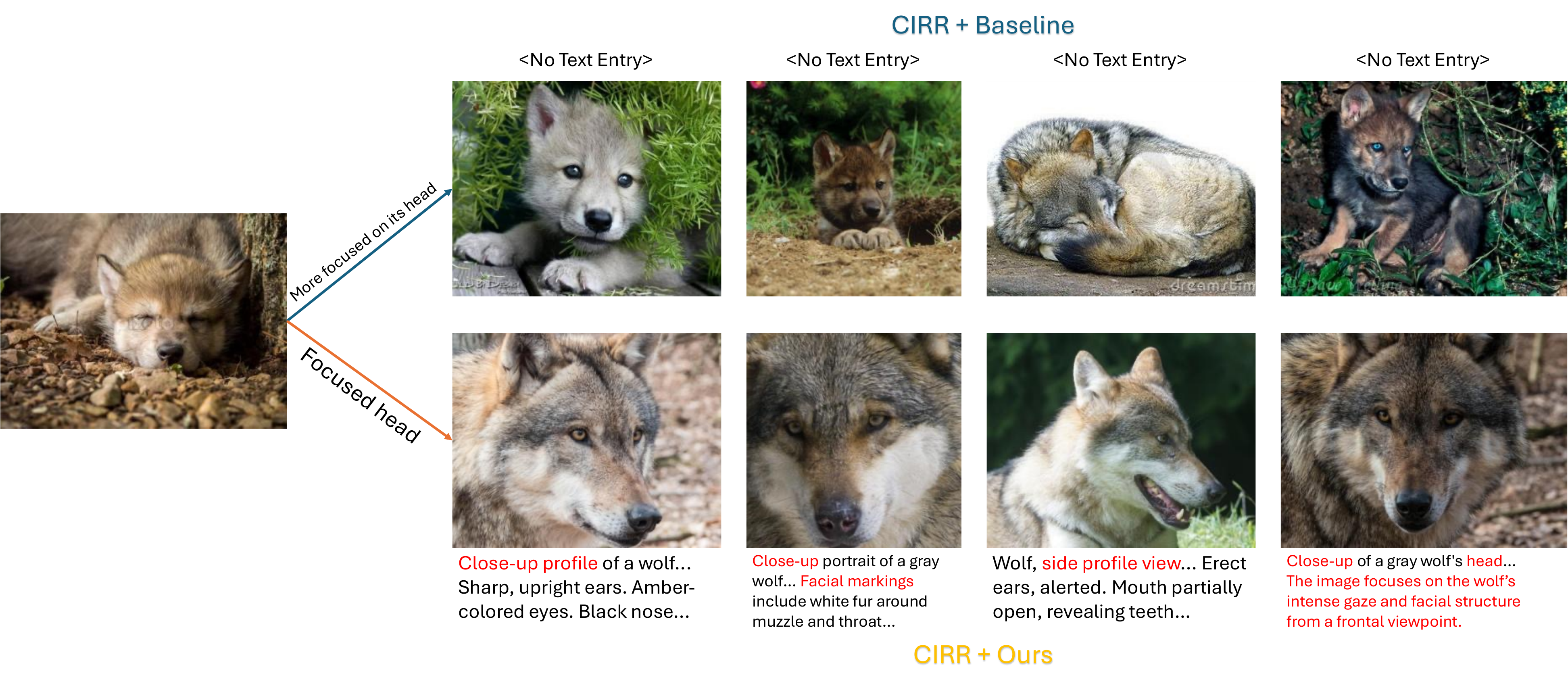}
    \vspace{-1em}
    \caption{Qualitative comparison on CIRR-7. The user's instruction requests a modification of ``more focus on its head.'' The baseline model, relying on implicit visual features, exhibits a strong bias toward low-level visual similarity, retrieving images with matching poses (sleeping bodies) rather than the requested semantic change. In contrast, our enhanced model leverages dense corpus captions (e.g., ``close-up,'' ``profile view'') to successfully align the modification instruction with the correct target.}
    \label{fig:qual_1}
    \includegraphics[width=.95\linewidth]{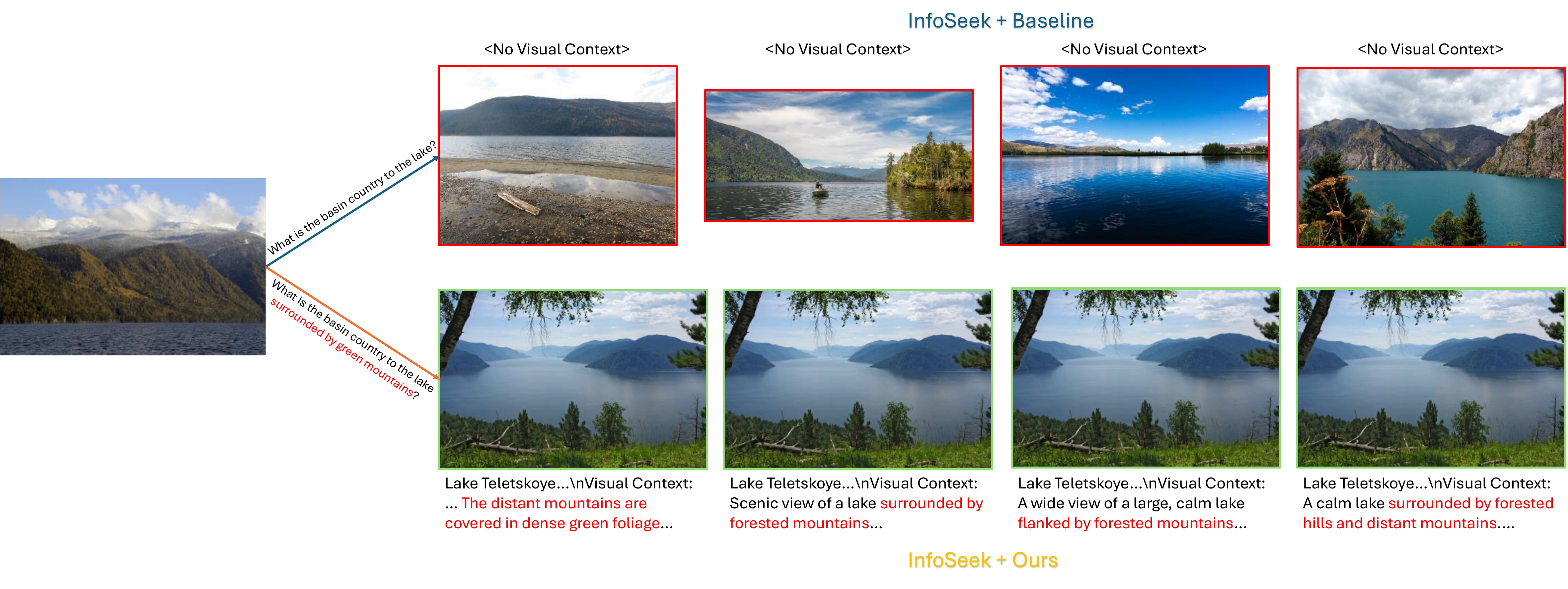}
    \vspace{-1em}
    \caption{Qualitative comparison on InfoSeek-8. The baseline struggles with the ambiguity of the multimodal query, where the visual signal alone is insufficient to anchor the vague question about ``the lake,'' leading to unrelated retrieval results. Our method bridges this gap by explicating visual cues; the generated context ``surrounded by green mountains'' in both the query and corpus serves as a crucial semantic bridge, enabling the precise retrieval of the correct Wikipedia entries.}
    \label{fig:qual_2}
\end{figure*}

\subsection{Qualitative Analysis}
\label{sec: qual}

We present qualitative examples in Figures~\ref{fig:qual_1} and~\ref{fig:qual_2} in the Appendix to demonstrate the efficacy of our reasoning-augmented framework compared to the baseline.

In Figure~\ref{fig:qual_1}, we analyze a query from \textit{CIRR-7} where the user instructs the system to retrieve an image that focuses on the wolf's head. As the corpus consists of image-only entries, the baseline model is forced to implicitly reason over the visual concept of ``focus.'' Consequently, it fails to capture the semantic shift, defaulting to spurious correlations by retrieving images of puppies in the same sleeping pose as the reference image. Conversely, our method enriches the corpus with detailed descriptions. Keywords such as ``close-up,'' ``profile view,'' and ``facial features'' in the enhanced captions provide a direct semantic match for the simplified query ``focused head,'' enabling the model to retrieve the correct target with high precision.

Figure~\ref{fig:qual_2} illustrates a challenging example from \textit{InfoSeek-8}. The query asks a specific question about a lake but relies on a generic image without explicit textual naming. Since the original Wikipedia entries lack detailed descriptions of their accompanying visual data, the baseline model fails to ground the query ``this lake'' to a specific entity, returning irrelevant lake-related articles. However, our pipeline injects dense visual cues into both the query and the corpus. In this instance, the explicit descriptor ``surrounded by green mountains'' was generated for both the query image and the target corpus images. This shared textual context acts as a robust retrieval anchor, allowing the model to filter out generic matches and correctly identify the specific location.

Collectively, these qualitative examples highlight a fundamental shift in the retrieval mechanism. The baseline model frequently overfits to \textbf{spurious visual correlations}—matching the sleeping pose of the wolf or the general color palette of the lake landscape—rather than capturing the user's intent. By incorporating these implicit visual attributes into text, our method allows the model to not conduct visual reasoning. This confirms that the performance gains reported in \S~\ref{sec: main results} are driven by the model's improved ability to ground abstract concepts (e.g., ``focus,'' ``this lake'') in concrete, textual definitions, thereby bypassing the ambiguity inherent in visual feature-level matching.
\section{Conclusion}
In this work, we argue that the key bottleneck in universal multimodal retrieval is not model capacity but semantic ambiguity in raw multimodal inputs, which forces retrievers to collapse \emph{reasoning} and \emph{compression} into a single embedding pass and encourages spurious matching. We introduce a data-centric framework that uses a strong VLM to \emph{externalize} these latent reasoning steps, turning ``reasoning-then-retrieve'' into explicit semantic matching over reasoning-augmented text. On M-BEIR, this yields consistent, substantial gains across tasks ranging from knowledge-intensive entity linking to fine-grained composed image modification. We further show that inference-time enhancement alone is insufficient: models must be trained on the dense semantic distribution induced by augmentation to avoid distribution shift and fully benefit from explicit semantics. Overall, our results suggest a practical route to more robust and interpretable multimodal retrieval by prioritizing scalable semantic augmentation and alignment over architectural complexity.




\section*{Impact Statement} This paper presents work whose goal is to advance the field of Machine Learning. There are many potential societal consequences of our work, none which we feel must be specifically highlighted here.

\section*{Acknowledgments}

This work was supported in part by NetApp Inc., NSF IIS2404180, and the Institute of Information \& Communications Technology Planning \& Evaluation (IITP) grants funded by the Korea government (MSIT) (No.RS-2025-02219317, AI Star Fellowship (Kookmin University)), (No. 2022-0-00871, Development of AI Autonomy and Knowledge Enhancement for AI Agent Collaboration), (No. RS-2022-00187238, Development of Large Korean Language Model Technology for Efficient Pretraining), and (No. RS-2025-2543949. Environment-Aware and Domain-Adaptive Multimodal Embodied AI for Real-World Interaction).

\bibliography{example_paper}

@misc{bai2025qwen3vltechnicalreport,
      title={Qwen3-VL Technical Report}, 
      author={Shuai Bai and Yuxuan Cai and Ruizhe Chen and Keqin Chen and Xionghui Chen and Zesen Cheng and Lianghao Deng and Wei Ding and Chang Gao and Chunjiang Ge and Wenbin Ge and Zhifang Guo and Qidong Huang and Jie Huang and Fei Huang and Binyuan Hui and Shutong Jiang and Zhaohai Li and Mingsheng Li and Mei Li and Kaixin Li and Zicheng Lin and Junyang Lin and Xuejing Liu and Jiawei Liu and Chenglong Liu and Yang Liu and Dayiheng Liu and Shixuan Liu and Dunjie Lu and Ruilin Luo and Chenxu Lv and Rui Men and Lingchen Meng and Xuancheng Ren and Xingzhang Ren and Sibo Song and Yuchong Sun and Jun Tang and Jianhong Tu and Jianqiang Wan and Peng Wang and Pengfei Wang and Qiuyue Wang and Yuxuan Wang and Tianbao Xie and Yiheng Xu and Haiyang Xu and Jin Xu and Zhibo Yang and Mingkun Yang and Jianxin Yang and An Yang and Bowen Yu and Fei Zhang and Hang Zhang and Xi Zhang and Bo Zheng and Humen Zhong and Jingren Zhou and Fan Zhou and Jing Zhou and Yuanzhi Zhu and Ke Zhu},
      year={2025},
      eprint={2511.21631},
      archivePrefix={arXiv},
      primaryClass={cs.CV},
      url={https://arxiv.org/abs/2511.21631}, 
}

@inproceedings{liu2024lamralargemultimodalmodel,
      title={LamRA: Large Multimodal Model as Your Advanced Retrieval Assistant}, 
      author={Yikun Liu and Pingan Chen and Jiayin Cai and Xiaolong Jiang and Yao Hu and Jiangchao Yao and Yanfeng Wang and Weidi Xie},
      year={2025},
      booktitle={CVPR} 
}

@misc{jiang2024e5vuniversalembeddingsmultimodal,
      title={E5-V: Universal Embeddings with Multimodal Large Language Models}, 
      author={Ting Jiang and Minghui Song and Zihan Zhang and Haizhen Huang and Weiwei Deng and Feng Sun and Qi Zhang and Deqing Wang and Fuzhen Zhuang},
      year={2024},
      eprint={2407.12580},
      archivePrefix={arXiv},
      primaryClass={cs.CL},
      url={https://arxiv.org/abs/2407.12580}, 
}

@inproceedings{jiang2025vlm2vectrainingvisionlanguagemodels,
      title={VLM2Vec: Training Vision-Language Models for Massive Multimodal Embedding Tasks}, 
      author={Ziyan Jiang and Rui Meng and Xinyi Yang and Semih Yavuz and Yingbo Zhou and Wenhu Chen},
      year={2025},
      booktitle={ICLR}
}

@misc{li2025umarvelunveilingkeyfactors,
      title={U-MARVEL: Unveiling Key Factors for Universal Multimodal Retrieval via Embedding Learning with MLLMs}, 
      author={Xiaojie Li and Chu Li and Shi-Zhe Chen and Xi Chen},
      year={2025},
      eprint={2507.14902},
      archivePrefix={arXiv},
      primaryClass={cs.IR},
      url={https://arxiv.org/abs/2507.14902}, 
}

@inproceedings{faysse2025colpaliefficientdocumentretrieval,
      title={ColPali: Efficient Document Retrieval with Vision Language Models}, 
      author={Manuel Faysse and Hugues Sibille and Tony Wu and Bilel Omrani and Gautier Viaud and Céline Hudelot and Pierre Colombo},
      year={2025},
      booktitle={ICLR}
}

@inproceedings{zhang2025gmeimprovinguniversalmultimodal,
      title={GME: Improving Universal Multimodal Retrieval by Multimodal LLMs}, 
      author={Xin Zhang and Yanzhao Zhang and Wen Xie and Mingxin Li and Ziqi Dai and Dingkun Long and Pengjun Xie and Meishan Zhang and Wenjie Li and Min Zhang},
      year={2025},
      booktitle={CVPR}
}

@misc{meng2025vlm2vecv2advancingmultimodalembedding,
      title={VLM2Vec-V2: Advancing Multimodal Embedding for Videos, Images, and Visual Documents}, 
      author={Rui Meng and Ziyan Jiang and Ye Liu and Mingyi Su and Xinyi Yang and Yuepeng Fu and Can Qin and Zeyuan Chen and Ran Xu and Caiming Xiong and Yingbo Zhou and Wenhu Chen and Semih Yavuz},
      year={2025},
      eprint={2507.04590},
      archivePrefix={arXiv},
      primaryClass={cs.CV},
      url={https://arxiv.org/abs/2507.04590}, 
}

@misc{li2026qwen3vlembeddingqwen3vlrerankerunifiedframework,
      title={Qwen3-VL-Embedding and Qwen3-VL-Reranker: A Unified Framework for State-of-the-Art Multimodal Retrieval and Ranking}, 
      author={Mingxin Li and Yanzhao Zhang and Dingkun Long and Keqin Chen and Sibo Song and Shuai Bai and Zhibo Yang and Pengjun Xie and An Yang and Dayiheng Liu and Jingren Zhou and Junyang Lin},
      year={2026},
      eprint={2601.04720},
      archivePrefix={arXiv},
      primaryClass={cs.CL},
      url={https://arxiv.org/abs/2601.04720}, 
}

@inproceedings{radford2021learningtransferablevisualmodels,
      title={Learning Transferable Visual Models From Natural Language Supervision}, 
      author={Alec Radford and Jong Wook Kim and Chris Hallacy and Aditya Ramesh and Gabriel Goh and Sandhini Agarwal and Girish Sastry and Amanda Askell and Pamela Mishkin and Jack Clark and Gretchen Krueger and Ilya Sutskever},
      year={2021},
      booktitle={Proceedings of the 38th International Conference on Machine Learning, PMLR} 
}

@inproceedings{zhai2023sigmoidlosslanguageimage,
      title={Sigmoid Loss for Language Image Pre-Training}, 
      author={Xiaohua Zhai and Basil Mustafa and Alexander Kolesnikov and Lucas Beyer},
      year={2023},
      booktitle={International Conference on Computer Vision}
}

@inproceedings{wei2023uniirtrainingbenchmarkinguniversal,
  title={Uniir: Training and benchmarking universal multimodal information retrievers},
  author={Wei, Cong and Chen, Yang and Chen, Haonan and Hu, Hexiang and Zhang, Ge and Fu, Jie and Ritter, Alan and Chen, Wenhu},
  booktitle={European Conference on Computer Vision},
  pages={387--404},
  year={2024},
  organization={Springer}
}

@inproceedings{girdhar2023imagebind,
  title={ImageBind: One Embedding Space To Bind Them All},
  author={Girdhar, Rohit and El-Nouby, Alaaeldin and Liu, Zhuang
and Singh, Mannat and Alwala, Kalyan Vasudev and Joulin, Armand and Misra, Ishan},
  booktitle={CVPR},
  year={2023}
}

@article{liu2023visual,
  title={Visual instruction tuning},
  author={Liu, Haotian and Li, Chunyuan and Wu, Qingyang and Lee, Yong Jae},
  journal={Advances in neural information processing systems},
  volume={36},
  pages={34892--34916},
  year={2023}
}

@inproceedings{khattab2020colbert,
  title={Colbert: Efficient and effective passage search via contextualized late interaction over bert},
  author={Khattab, Omar and Zaharia, Matei},
  booktitle={Proceedings of the 43rd International ACM SIGIR conference on research and development in Information Retrieval},
  pages={39--48},
  year={2020}
}

@inproceedings{ma2024unifyingmultimodalretrievaldocument,
      title={Unifying Multimodal Retrieval via Document Screenshot Embedding}, 
      author={Xueguang Ma and Sheng-Chieh Lin and Minghan Li and Wenhu Chen and Jimmy Lin},
      year={2024},
      booktitle={EMNLP}
}

@inproceedings{liu2025any,
  title={Any information is just worth one single screenshot: Unifying search with visualized information retrieval},
  author={Liu, Zheng and Liu, Ze and Liang, Zhengyang and Zhou, Junjie and Xiao, Shitao and Gao, Chao and Zhang, Chen Jason and Lian, Defu},
  booktitle={Proceedings of the 63rd Annual Meeting of the Association for Computational Linguistics (Volume 1: Long Papers)},
  pages={19238--19261},
  year={2025}
}

@article{ma2024unifying,
  title={Unifying multimodal retrieval via document screenshot embedding},
  author={Ma, Xueguang and Lin, Sheng-Chieh and Li, Minghan and Chen, Wenhu and Lin, Jimmy},
  journal={arXiv preprint arXiv:2406.11251},
  year={2024}
}
\bibliographystyle{icml2026}

\newpage
\appendix
\onecolumn

\section{Prompts Used}
\label{app: prompts}

\subsection{Corpus Enhancement}
The prompt we feed to Qwen for enhancing corpus entries is as follows:

\noindent \textbf{Task:} Generate a precise, keyword-rich text entry based on the [Image].

\noindent \textbf{Instructions:}
\begin{enumerate} 
    \item \textbf{Subject First:} Identify the main object, entity, or scene layout immediately.
    \item \textbf{Distinctive Features:} List specific details: colors, materials, text/logos (if visible), and unique shapes. If a detail doesn't exist, don't mention it (no stating `no visible logos or text').
    \item \textbf{Entity Recognition:} If the object is a named entity (e.g., `Eiffel Tower', `Toyota Camry', `Nike'), state it.
    \item \textbf{Viewpoint:} Mention the angle (e.g., `close-up', `aerial', `profile') ONLY IF it distinguishes the image.
    \item \textbf{No Filler:} Do not use aesthetic words (e.g., `beautiful', `cinematic'). Focus on factual visual content.
    \item \textbf{Length:} Maximum 100 words.
\end{enumerate}

\noindent \textbf{Reference Image:} \texttt{<image>} \\
\noindent \textbf{Output:}

\subsection{QA Query Enhancement}

The prompt we feed to Qwen for enhancing QA format queries is as follows:

\noindent \textbf{Task:} Rewrite the user's question by integrating the visual subject. \\
\textbf{Goal:} Create a search query that matches text documents. Keep it extremely concise.

\textbf{Strict Constraint Rules:}

\begin{enumerate}
    \item \textbf{Length Limit:} The added visual description must be MAX 3--5 words. No long sentences.
    
    \item \textbf{The `Specific vs. Generic' Split:}
    \begin{itemize}
        \item \textbf{If Unique Entity (Landmark, Art, Car Model):} Use the NAME only. Delete all visual adjectives.
        \begin{itemize}
            \item BAD: `Who built this tall iron tower?'
            \item GOOD: `Who built the Eiffel Tower?'
        \end{itemize}
        \item \textbf{If Generic Object (Food, Plant, Animal):} Use [Dominant Color/Material] + [Broad Category].
        \begin{itemize}
            \item BAD: `What is this delicious spicy red soup with shrimp?' (Too many distractors)
            \item GOOD: `What is this red noodle soup with shrimp?' (Anchors only)
        \end{itemize}
    \end{itemize}

    \item \textbf{No `Filler' Adjectives:} Banned words: `beautiful', `large', `small', `generic', `distinct', `looking', `shaped'.
    
    \item \textbf{No Environment:} Never mention background, weather, or lighting.
    
    \item \textbf{Zero-Leakage:} NEVER answer the question yourself. YOU ARE ONLY REWRITING THE QUERY.
\end{enumerate}

\textbf{Examples:}

\begin{description}
    \item[Input:] [Photo of Giant Panda] $|$ \textbf{Query:} `When was it discovered?'
    \item[Output:] When was the Giant Panda discovered? \\
    (\textit{Reason: Named entity. No adjectives needed.})
    
    \vspace{0.5em}

    \item[Input:] [Photo of Yellowjacket Wasp] $|$ \textbf{Query:} `What species is this?'
    \item[Output:] What species is this black and yellow wasp? \\
    (\textit{Reason: `Black and yellow' distinguishes it. `Insect' is too broad, `Wasp' is better.})

    \vspace{0.5em}

    \item[Input:] [Photo of Red Laksa Soup] $|$ \textbf{Query:} `What dish is this?'
    \item[Output:] What dish is this red noodle soup with shrimp? \\
    (\textit{Reason: `Red', `Noodle', `Shrimp' are the only keys needed to find the recipe.})

    \vspace{0.5em}

    \item[Input:] [Photo of Blue Ford Focus] $|$ \textbf{Query:} `What car is this?'
    \item[Output:] What car is this blue hatchback? \\
    (\textit{Reason: `Blue' and `Hatchback' filter the candidates. `Ford Focus' might be a hallucination, so we play it safe. We also don't want to leak the answer.})

    \vspace{0.5em}

    \item[Input:] [Photo of Melting Clock Painting] $|$ \textbf{Query:} `Who painted this?'
    \item[Output:] Who painted The Persistence of Memory? \\
    (\textit{Reason: Unique Art $\to$ Specific Name.})
\end{description}

\noindent \textbf{Current Task:} \\
\textbf{Query:} \texttt{\{query\_txt\}} \\
\textbf{Input Image:} \texttt{<image>} \\
\noindent \textbf{Output:}

\subsection{Modification Request Enhancement}

The prompt we feed to Qwen for enhancing modification requests is as follows:

\noindent \textbf{Task:} Extract the key semantic phrases describing the TARGET image. Remove conversational filler and grammar words. \\
\textbf{Input:} User Query (describing a change or a target attribute).

\textbf{Strict Reduction Rules:}

\begin{enumerate}
    \item \textbf{Delete Filler Verbs:} Remove `Is', `Has', `Make', `Change', `Show', `Put', `Be'.
    \item \textbf{Delete Pronouns/Articles:} Remove `it', `the', `a', `an', `my', `me', `them', `this'.
    \item \textbf{Preserve Adjectives \& Nouns:} Keep ALL descriptors (colors, patterns, objects). If the user says `Is white', output `White'.
    \item \textbf{Preserve Prepositions:} Keep `with', `on', `in', `without' to maintain spatial/compositional logic.
\end{enumerate}

\noindent \textit{Note: There are cases where the original query is concise enough, and you might not have to change anything.}

\textbf{Examples:}

\begin{description}
    \item[Input:] `Is shiny and silver with shorter sleeves.'
    \item[Output:] Shiny silver with shorter sleeves
    
    \vspace{0.5em}

    \item[Input:] `Is white in color with short sleeves and is more plain.'
    \item[Output:] White, short sleeves, more plain
    
    \vspace{0.5em}

    \item[Input:] `Remove the lemon.'
    \item[Output:] Remove lemon
    
    \vspace{0.5em}

    \item[Input:] `Make the needle upside down in the hand.'
    \item[Output:] Needle upside down in hand
    
    \vspace{0.5em}

    \item[Input:] `Human and one animal from a different species.'
    \item[Output:] Human and animal from different species
    
    \vspace{0.5em}

    \item[Input:] `Is a plain white feminine t shirt and is a tan shirt.'
    \item[Output:] Plain white feminine t-shirt and tan shirt
    
    \vspace{0.5em}

    \item[Input:] `Remove all cheetahs.'
    \item[Output:] Remove all cheetahs
    
    \vspace{0.5em}

    \item[Input:] `Remove one cheetah.'
    \item[Output:] Remove one cheetah.
    
    \vspace{0.5em}

    \item[Input:] `Remove green from the background.'
    \item[Output:] Remove green from background.
\end{description}

\noindent \textbf{Current Input:} \texttt{\{query\_txt\}}\\
\noindent \textbf{Output:}

\section{Experiments on MVRB Composed Image Retrieval}\label{app:mvrb}
\begin{table*}[t]
\centering
\caption{Baseline vs Our method on MVRB Composed Image Retrieval. Our method outperforms the Baseline model on the Composed Image Retrieval Section of the MVRB benchmark which involve complex queries that requires reasoning.}
\label{tab:knowledge_recall}
\begin{tabular}{l|cc|cc}
\toprule
\multirow{2}{*}{\textbf{Task}} 
& \multicolumn{2}{c|}{\textbf{Baseline \tiny{(LamRA-Ret~\citep{liu2024lamralargemultimodalmodel})}}} 
& \multicolumn{2}{c}{\textbf{Ours}} \\
& \textbf{R@1} & \textbf{R@5} 
& \textbf{R@1} & \textbf{R@5} \\
\midrule
Knowledge Relation & 35.00 & 71.00 & \textbf{46.00} & \textbf{78.00} \\
News to Wiki       & 52.48 & \textbf{85.15} & \textbf{55.45} & 83.17 \\
Product Discovery  & \textbf{52.34} & 91.59 & 50.47 & \textbf{92.52} \\
Wiki to Product    & 70.00 & \textbf{97.00} & \textbf{73.00} & 96.00 \\
\bottomrule
\end{tabular}
\end{table*}

We further evaluate our approach on composed image retrieval tasks from the MVRB benchmark \citep{liu2025any}. These tasks are particularly challenging, as they require multi-step reasoning over object relations (e.g., identifying the phone case corresponding to a given phone). As shown in Fig.~\ref{tab:knowledge_recall}, our method consistently outperforms the baseline on several tasks as indicated by the recall scores. The performance gains are especially pronounced on knowledge-based retrieval tasks, further supporting the claims made in Section~\ref{baseline-comp}.

\section{Impact of Hard Negative Mining and Data Enrichment}
\label{app: hard negs}

We evaluate the effectiveness of our Enriched (Full) data when integrated into a robust Hard Negative training regime. Table~\ref{tab:hardnegs} compares the performance of the Baseline model versus our Full model, both trained with M-BEIR hard negatives. 

\begin{table*}
\centering
\caption{\textbf{Performance comparison of Baseline vs. Full data under Hard Negative training.} While the Full model demonstrates consistent improvements across the majority of tasks (see Avg w/o outliers), specific regressions are observed on OVEN-8 and EDIS-2. We attribute these drops to task-specific interference introduced by the hard negative mining process, as discussed in Section~\ref{sec:analysis}.}
\label{tab:hardnegs}
\begin{tabular}{l|l|cccc|cccc}
\toprule
\multirow{2}{*}{\textbf{Task}} & \multirow{2}{*}{\textbf{Type}} & \multicolumn{4}{c|}{\textbf{Baseline + Hard Negatives}} & \multicolumn{4}{c}{\textbf{Ours + Hard Negatives}} \\
 & & \textbf{R@1} & \textbf{R@5} & \textbf{R@10} & \textbf{R@50} & \textbf{R@1} & \textbf{R@5} & \textbf{R@10} & \textbf{R@50} \\
\midrule
MSCOCO-0 & $q^t\to c^i$ & 53.93 & 79.04 & 86.88 & 97.93 & \textbf{55.09} & \textbf{79.87} & \textbf{87.48} & \textbf{97.99} \\
MSCOCO-3 & $q^i\to c^t$ & 69.86 & 89.46 & 94.42 & 99.54 & \textbf{70.54} & \textbf{89.98} & \textbf{94.64} & 99.40 \\
VisNews-0 & $q^t\to c^i$ & 13.70 & 28.39 & 35.33 & 53.65 & \textbf{14.76} & \textbf{29.28} & \textbf{36.41} & \textbf{55.18} \\
VisNews-3 & $q^i\to c^t$ & 12.10 & 25.90 & 33.12 & 51.99 & \textbf{12.85} & \textbf{26.56} & \textbf{34.04} & \textbf{52.51} \\
WebQA-1 & $q^t\to c^t$ & 74.34 & 93.08 & 95.56 & 98.86 & 74.18 & \textbf{93.12} & \textbf{95.89} & 98.78 \\
WebQA-2 & $q^t\to c^i,c^t$ & 55.87 & 83.19 & 90.56 & 97.25 & \textbf{56.59} & \textbf{83.59} & \textbf{91.20} & \textbf{97.45} \\
InfoSeek-6 & $q^i,q^t\to c^t$ & 20.53 & 39.15 & 49.02 & 69.97 & \textbf{21.37} & \textbf{40.10} & 48.70 & 68.87 \\
InfoSeek-8 & $q^i,q^t\to c^i,c^t$ & 33.11 & 55.28 & 64.04 & 83.07 & \textbf{33.64} & \textbf{56.62} & \textbf{65.80} & 82.49 \\
OVEN-6 & $q^i,q^t\to c^t$ & 22.22 & 44.33 & 54.13 & 73.63 & \textbf{22.46} & 44.28 & 53.88 & 73.30 \\
OVEN-8 & $q^i,q^t\to c^i,c^t$ & 41.84 & 65.35 & 72.40 & 85.52 & 40.88 & 63.89 & 71.21 & 84.68 \\
EDIS-2 & $q^t\to c^i,c^t$ & 33.88 & 56.65 & 65.44 & 83.61 & 30.17 & 53.10 & 62.67 & 82.65 \\
Nights-4 & $q^i\to c^i$ & 9.34 & 32.50 & 53.44 & 90.94 & \textbf{9.48} & \textbf{33.35} & \textbf{54.20} & 90.80 \\
CIRR-7 & $q^i\to c^i$ & 23.09 & 50.10 & 61.29 & 82.78 & \textbf{23.65} & 49.81 & 59.93 & 82.09 \\
\color{gray} FashIQ-7 & \color{gray} $q^i,q^t\to c^i$ & \color{gray} 10.48 & \color{gray} 23.06 & \color{gray} 30.38 & \color{gray} 48.59 & \color{gray} 10.31 & \color{gray} \textbf{23.49} & \color{gray} 30.25 & \color{gray} \textbf{49.06} \\
\color{gray} Fash200k-3 & \color{gray} $q^i\to c^t$ & \color{gray} 4.93 & \color{gray} 13.25 & \color{gray} 18.70 & \color{gray} 37.35 & \color{gray} \textbf{4.95} & \color{gray} 12.62 & \color{gray} 18.08 & \color{gray} 36.78 \\
\bottomrule
\end{tabular}%
\end{table*}

\subsection{Robustness of Semantic Enrichment}
The primary objective of this experiment was to stress-test our Enriched (Full) data against a highly competitive baseline strengthened by hard negative mining. Despite the significantly elevated performance bar set by the hard-negative baseline, our Full model successfully provides additive performance gains across the majority of benchmarks. Notably, our findings of improvement trends highly match our findings in the main experiments, confirming that the benefits of semantic enrichment are robust and persist even under rigorous discriminative training regimes.

On general retrieval benchmarks such as MSCOCO, we observe clear improvements (e.g., +1.16\% R@1 on MSCOCO-0), reaffirming that detailed captions enhance fundamental scene understanding. This positive trend extends to knowledge-intensive and multimodal tasks like VisualNews and InfoSeek, where our method consistently outperforms the baseline. The gains on InfoSeek-6 and InfoSeek-8 are particularly significant, as they demonstrate that our enriched entity descriptions and factual augmentations provide critical discriminative signals. These signals allow the model to resolve complex queries more effectively than negative mining alone, proving that our data enrichment strategy is not merely redundant but complementary to the structural improvements gained from hard negative training.

\subsection{Analysis of Task-Specific Deviations}
\label{sec:analysis}
While the general trend is positive, we observe specific regressions on OVEN-8 and EDIS-2. We attribute these deviations to \textbf{dataset-specific interference} and \textbf{bias introduction} during the hard negative mining process.

\paragraph{OVEN-8 and Cross-Task Interference}
The performance regression on OVEN-8 is most accurately understood by first examining the Baseline's behavior. The introduction of hard negatives caused a significant $\sim$6 point drop in the Baseline performance compared to the original training configuration. This explicitly isolates the source of the regression to biases induced by the hard negative training regime itself, rather than an intrinsic flaw in our data enrichment process.

OVEN serves as a coarser, less nuanced counterpart to InfoSeek, sharing the same underlying data distribution. However, while InfoSeek relies on explicit hard negatives to enforce fine-grained discrimination, OVEN's training data does not. We hypothesize that training on the shared InfoSeek hard negatives introduces a distribution bias that conflicts with the objectives of OVEN. The model likely overfits to the hyper-specific discriminative features required for InfoSeek, which do not generalize to---and actively interfere with---the broader entity retrieval requirements of OVEN-8.

\paragraph{EDIS-2 and Length-Induced Modality Bias}
The regression on EDIS-2 reveals a critical interaction between data length and hard negative training. EDIS is characterized by concise text corpus entries, significantly shorter than the encyclopedic descriptions found in benchmarks like InfoSeek. However, our enrichment pipeline uniformly generates detailed captions ($\sim$100 words) regardless of the native corpus density. 

Crucially, in our prior experiments without hard negatives, our Full model outperformed the Baseline on EDIS-2, confirming that our generated data is intrinsically high-quality and informative. The performance reversal occurs only under hard negative training. We hypothesize that the stark contrast between our verbose generated captions and the concise ground truth creates a \textbf{modality imbalance}. When forced to discriminate against hard negatives, the model may over-index on the abundant visual context provided by our long descriptions, while demonstrating diminished sensitivity to the sparse but critical textual cues present in the short ground truth. This suggests that while semantic expansion is beneficial generally, rigid length constraints can introduce signal-to-noise issues in low-density text domains.

\paragraph{Future Work: Dynamic Length Curation}
To resolve the length mismatch observed in EDIS and the interference in OVEN, our future work will pivot from fixed-length generation to \textbf{dynamic length curation}. Rather than systematically generating 100-word descriptions, we aim to condition the generation pipeline to match the intrinsic information density and length distribution of the target domain. By adaptively sizing the enriched text—providing verbose descriptions for complex scenes (e.g., InfoSeek) and concise, sharp summaries for sparse domains (e.g., EDIS)—we aim to retain the semantic benefits of enrichment while preventing the distribution shifts that currently hamper fine-grained discrimination.

\end{document}